\documentstyle[12pt,fleqn,epsf]{article}
\def\nsection#1{\setcounter{equation}{0}\section{#1}}

\newenvironment{eqar}{
	\setlength{\mathindent}{0 cm}\begin{eqnarray}}{\end{eqnarray}}
\newenvironment{eq}{
	\setlength{\mathindent}{0 cm}\begin{equation}}{\end{equation}}
\newcommand{\Integer}{\:\mbox{\sf Z} \hspace{-0.82em} \mbox{\sf Z}\,}
\def\C{\rule[0.5pt]{0.2mm}{7pt}{\hspace{-3.6pt}{\rm C}}}
\def\Mult#1#2#3{\left[#1 \atop #2 \right]_{#3}}
\def\Mults#1#2#3{\left[{\textstyle {#1 \atop #2} } \right]_{#3}}
\def\ib{\,\mbox{i}\,}
\def\is{\,\mbox{\scriptsize i}\,}
\def\e{\mbox{e}}
\def\d#1{\mbox{d}#1}
\def\case#1#2{{\textstyle{#1\over #2}}}
\def\rG{\mbox{\scriptsize rank}\, {\cal G}}
\def\A{\,\mbox{\scriptsize A}_{\ell-1}}
\def\An{\,\mbox{\scriptsize A}_{n-1}}
\def\Dn{\,\mbox{\scriptsize D}_{n-1}}
\def\CG{C^{\,\cal G}}
\def\m#1#2{m_{#1}^{(#2)}}
\def\n#1#2{n_{#1}^{(#2)}}
\def\ppmod#1{\!\!\! \pmod{#1}}

\begin{document}

\title{A-D-E Polynomial and Rogers--Ramanujan Identities}

\author{}
\author{S.~Ole Warnaar\thanks{
e-mail: {\tt warnaar@mundoe.maths.mu.oz.au}}
\ and
Paul A.~Pearce\thanks{
e-mail: {\tt  pap@mundoe.maths.mu.oz.au}} \\
Mathematics Department\\
University of Melbourne\\
Parkville, Victoria 3052\\
Australia}

\date{}

\maketitle

\begin{abstract}
We conjecture polynomial identities which imply Rogers--Ramanujan
type identities for branching functions  associated with
the cosets $({\cal G}^{(1)})_{\ell-1}\otimes ({\cal G}^{(1)})_{1} /
({\cal G}^{(1)})_{\ell}$, with
${\cal G}$=A$_{n-1}$ \mbox{$(\ell\geq 2)$}, D$_{n-1}$ $(\ell\geq 2)$,
E$_{6,7,8}$ $(\ell=2)$.
In support of our conjectures we establish the correct behaviour
under level-rank duality for $\cal G$=A$_{n-1}$ and show that the
A-D-E Rogers--Ramanujan identities have the expected $q\to 1^{-}$
asymptotics in terms of dilogarithm identities.
Possible generalizations to arbitrary cosets are also discussed briefly.
\end{abstract}

\nsection{Introduction}
Without doubt, the Rogers--Ramanujan identities
\begin{equation}
\sum_{m=0}^{\infty} \frac{q^{m(m+\sigma)}}{(q)_m} =
\frac{1}{(q)_{\infty}}
\sum_{j=-\infty}^{\infty} \left\{
q^{10j^2+(1-4\sigma)j}-
q^{10j^2+(11-4\sigma)j+3-2\sigma} \right\} \qquad \sigma=0,1,
\label{RogersRam}
\end{equation}
with $(q)_m = \prod_{k=1}^m (1-q^k)$ for $m>0$ and $(q)_0=1$,
are among the most beautiful and intriguing results
of classical mathematics.
Since their independent discovery by Rogers~\cite{Rogers}
and Ramanujan~\cite{Ramanujan} 
many different methods of proof have been developed.
A particularly fruitful approach
was initiated by Schur~\cite{Schur}.
The key idea is that identities of the Rogers--Ramanujan
type are in fact limiting
cases of polynomial identities.
For example, the polynomial identities
\begin{eqnarray}
\lefteqn{
\sum_{m=0}^{\infty} q^{m(m+\sigma)}
\Mult{L-m-\sigma}{m}{q} } \nonumber \\
& & \nonumber \\
& &\quad=
\sum_{j=-\infty}^{\infty} \left\{
q^{10j^2+(1-4\sigma)j}
\Mult{L}{\lfloor\frac{L}{2}\rfloor-5j+\sigma}{q}-
q^{10j^2+(11-4\sigma)j+3-2\sigma}
\Mult{L}{\lfloor\frac{L-5}{2}\rfloor-5j+\sigma}{q}
\right\},
\label{finiteRR}
\end{eqnarray}
which hold for arbitrary $L\in \Integer_{\geq 0}$,
yield the Rogers--Ramanujan identities~(\ref{RogersRam})
in the limit $L\to\infty$ when $q$ is restricted to $|q|<1$.
Here $\lfloor x \rfloor$ denotes the largest integer not exceeding $x$,
and the Gaussian- or $q$ polynomial is defined by \cite{Andrews}
\begin{equation}
\Mult{N}{m}{q} = \left\{
\begin{array}{ll}
\displaystyle \frac{(q)_N}{(q)_{m}(q)_{N-m}} \qquad & 0\leq m \leq N \\
& \\
0& \mbox{otherwise.}
\end{array}
\right.
\label{qpoly}
\end{equation}

Using the elementary formulae
$\Mults{N}{m}{q}=\Mults{N-1}{m}{q}+q^{N-m}\Mults{N-1}{m-1}{q}$ and
$\Mults{N}{m}{q}=\Mults{N-1}{m-1}{q}+q^{m}\Mults{N-1}{m}{q}$
it is easy to verify that
both the left- and right-hand-sides of
equation (\ref{finiteRR}) satisfy the recurrence
\mbox{$f_L=f_{L-1}+q^{L-1} f_{L-2}$}.
Given appropriate initial conditions,
this recurrence has a unique solution hence establishing
the polynomial identity and
its limiting form (\ref{RogersRam}).

\vspace*{5mm}

Identities of the Rogers--Ramanujan type
occur in various branches of mathematics and physics.
First, their connection with the theory of (affine)
Lie algebras~\cite{Kac,KP}
and with
partition theory~\cite{Andrews}
has led to many generalizations of (\ref{RogersRam}).
To illustrate these connections, it is
for example easily established that for $\sigma=1$ the
right-hand-side of (\ref{RogersRam}) can be rewritten in a more
algebraic fashion as\footnote{If a symbol $x$ is used in the
context of both classical and affine Lie algebras, we will
throughout this paper
write $\overline{x}$ to mean its classical (counter)part.}
\begin{equation}
\frac{q^{c/24}}{\eta(q)}
\sum_{\alpha \in Q} \, \sum_{w\in \overline{W}} \mbox{sgn}(w) \; \;
q^{\displaystyle
\case{1}{2} \, p \, p' \left| \alpha -\frac{p' \overline{\rho}
- p \, w(\overline{\rho})}{p\, p'}\right|^2}
\qquad \qquad p=2,p'=5,
\label{Virchar}
\end{equation}
with $Q$ the root lattice,
$\overline{W}$ the Weyl group and $\overline{\rho}$
the Weyl vector of the classical
Lie algebra A$_1$,
$\eta(q)= q^{1/24} (q)_{\infty}$
the Dedekind eta function and
\begin{equation}
c=1-\frac{6(p-p')^2}{p\, p'}.
\label{cc}
\end{equation}
Similarly, if $Q_{k,i}(n)$ denotes the number of partitions
of $n$ with each successive rank in the interval
$[2-i,2k-i-1]$, then by sieving methods the
generating function of $Q_{k,i}$ can be seen to
again yield the right-hand-side of (\ref{RogersRam})
provided we choose $k=2$ and $i=3-2\sigma$ \cite{Andrews}.

Second, Rogers--Ramanujan identities also appear
in various areas of physics.
Most notable is perhaps the fact that (\ref{Virchar})
can be identified
as the normalized Rocha-Caridi form \cite{Rocha} for the identity character
$\chi_{1,1}^{(p,p')}$ of the Virasoro algebra.
Indeed, each pair of positive integers $p$, $p'$ with
$p$ and $p'$ coprime, labels a minimal
conformal field theory \cite{BPZ}
of {\em central charge} $c$ given by (\ref{cc}).
Another branch of physics where Rogers--Ramanujan type identities
have occurred is in the theory of solvable lattice models~\cite{Baxter}.
Among other works,
in refs.~\cite{ABF,FB} Andrews, Baxter and Forrester (ABF) encountered
generalized Rogers--Ramanujan identities in their corner transfer matrix
(CTM) calculation
of one-point functions of an infinite series of
restricted solid-on-solid (RSOS) models.
In addition, in refs.~\cite{KKMMa,KKMMb}
Kedem {\em et al.}
conjectured many identities motivated by a Bethe Ansatz study
of the row transfer matrix spectrum of the 3-state Potts model.

Interestingly though, it is in fact the combination of the
approaches of ref.~\cite{ABF} and ref.~\cite{KKMMa} to solvable
models that leads to polynomial identities of the
type (\ref{finiteRR}).
In computing one-point functions of solvable RSOS models
using CTMs along the lines of ref.~\cite{ABF} one is naturally
led to the computation of so-called one-dimensional
configuration sums. These configuration sums take forms
very similar to the right-hand-side of (\ref{finiteRR}).
On the other hand, in performing Bethe Ansatz, and more
particularly Thermodynamic Bethe Ansatz (TBA)
computations, one is led to expressions of a similar nature
to the left-hand-side of (\ref{finiteRR}).

Starting with the ABF models and pursuing
the lines sketched above,
Melzer conjectured \cite{Melzer}
an infinite family of polynomial identities similar
to those in (\ref{finiteRR}).
In the infinite limit these identities again lead to Rogers--Ramanujan
type identities, but now for Virasoro characters
$\chi_{r,s}^{(p,p+1)}$ of unitary minimal
models, i.e., for characters with the Rocha-Caridi right-hand-side
form (in the sense of (\ref{RogersRam}))
\begin{equation}
\chi_{r,s}^{(p,p+1)}(q) = \frac{q^{\Delta_{r,s}^{(p,p+1)}-c/24}}{(q)_{\infty}}
\sum_{j=-\infty}^{\infty}
\left\{
q^{p(p+1)j^2+[(p+1)r-ps]j}-
q^{p(p+1)j^2+[(p+1)r+ps]j+rs} \right\}
\label{Xrs}
\end{equation}
labelled by the conformal weights
\begin{equation}
\Delta_{r,s}^{(p,p+1)} = \frac{[(p+1)r-ps]^2-1}{4p(p+1)} \quad
\qquad 1\leq r \leq p-1, \quad 1\leq s \leq p.
\label{weights}
\end{equation}
The corresponding left-hand-side forms of  these characters were
conjectured earlier in the work
of Kedem {\em et al.} \cite{KKMMb} and their {\em finitization}
in \cite{Melzer}
again provided a method of proof. For $p=3$ and 4 the proof
was carried out in ref.~\cite{Melzer} and Berkovich \cite{Berkovich}
subsequently generalized this to all $p\geq 3$ for $s=1$.

In this paper we generalize Melzer's approach to
finding polynomial identities from solvable lattice models.
By considering the CTM
as well as TBA calculations of various higher rank generalizations of the
ABF models, we are led to conjectures for polynomial identities labelled
by the Lie algebras of A,D and E type.
In the infinite limit our $\cal G$=A-D-E polynomial identities
lead to Rogers--Ramanujan type expressions for
the branching functions associated with the GKO coset pair \cite{GKO,Kac}
\begin{equation}
\begin{array}{rcccccc}
&{\cal G}^{(1)} &\oplus & {\cal G}^{(1)} &\supset & {\cal G}^{(1)} & \\
\mbox{level } &\ell-1& &  1& & \ell &   .
\label{GKO}
\end{array}
\end{equation}

The rest of this paper is organized as follows.
In the next section we define polynomial expressions
$F_q^{\,\cal G}(L)$
following from the TBA calculations of Bazhanov and Reshetikhin
\cite{BR}. Since the polynomials are defined as
restricted sums over the solutions of $\cal G$ type
constraint equations, we adopt the terminology of
ref.~\cite{KKMMa} and call these polynomials {\em fermionic}.
In section~\ref{sec3}, we define analogous polynomial expressions,
originating from the CTM calculations of
refs.~\cite{JMOb,DJKMO,WPSN}
for the same solvable lattice
models as considered by Bazhanov and Reshetikhin.
Again following the terminology of ref.~\cite{KKMMa},
we call these polynomials {\em bosonic}, and denote them
by $B_q^{\,\cal G}(L)$.
Our conjectures can then be formulated as
\begin{equation}
F_q^{\,\cal G}(L) =  B_q^{\,\cal G}(L).
\end{equation}
In section~\ref{sec4} we study the $L\to\infty$ behaviour of our
polynomial
identities, thereby deriving A-D-E type
Rogers--Ramanujan identities.
In the subsequent two sections we provide some
indications for the correctness of our conjectures.
In section~\ref{sec5} we establish the expected level-rank
duality for $\cal G$=A$_{n-1}$ and in section~\ref{sec6}
we show that in the $q\to 1^{-}$ limit
the A-D-E Rogers--Ramanujan identities
lead to the correct dilogarithm identities.
Finally, in section~\ref{sec7}, we summarize our results
and discuss possible generalizations
to non simply-laced Lie algebras and to more
general cosets than those listed under (\ref{GKO}).

\nsection{Fermionic A-D-E Polynomials}\label{sec2}
Let us now turn to the definition of the fermionic A-D-E
polynomials as follow from the TBA calculations
of ref.~\cite{BR}.

We denote the Cartan matrix of the
simply-laced Lie algebra ${\cal G}$=A$\,$,D,$\,$E
by $\CG$ and the corresponding
incidence matrix by ${\cal I}^{\, \cal G}=2\;  \mbox{Id} -\CG$,
choosing the labelling of the nodes of the
A-D-E Dynkin diagrams as shown in figure~1.
We furthermore
let $\m{j}{a}\in\Integer_{\geq 0}$ and
$\n{j}{a}\in\Integer_{\geq 0}$ be the number of `particles' and
`anti-particles' of type $j$ and colour $a$, respectively, satisfying the
following constraint system:
\begin{equation}
\m{j}{a} + \n{j}{a} =
\frac{1}{2} \left[ L \: \delta_{a,p} \delta_{j,1} +
\sum_{b=1}^{\rG}
{\cal I}^{\,\cal G}_{a,b}\:
\n{j}{b} +
\sum_{k=1}^{\ell-1} {\cal I}^{\A}_{j,k}
\m{k}{a} \right].
\label{TBAconstr}
\end{equation}
Here the particles are labelled from $j=1,\ldots,\ell-1$ and
the colours from
$a=1,\ldots,\mbox{rank}\, \cal G$. The variable $p$ in the
first Kronecker delta is 1 except for $\cal G$=E$_{6,7}$
when we have $p=6$.
We note that the above equation is a parameter independent
version of the constraint equations for densities of strings
and holes in the TBA
calculations of refs.~\cite{BR,Kuniba}, and given a set
$\{ m_j^{(a)} \}_{j=1,\ldots,\mbox{\scriptsize rank}\, \cal
G}^{a=1,\ldots, \ell-1}$ it determines
a companion set
$\{ n_j^{(a)} \}_{j=1,\ldots,\mbox{\scriptsize rank}\, \cal
G}^{a=1,\ldots, \ell-1}\,$.
Of course, only for
special sets $\{\m{j}{a}\}$ does it follow
that $\{\n{j}{a}\}$ again consists of only non-negative integers.

We now define the fermionic polynomials
$F_q^{\,\cal G}(L)$ as the following
sum over the solutions to the constraint system (\ref{TBAconstr}):
\begin{equation}
F_q^{\,\cal G}(L) = \left. \sum \right.^{'}q^{\; \displaystyle \case{1}{2}
\sum_{a,b=1}^{\rG} \sum_{j,k=1}^{\ell-1}
\left(\CG\right)^{-1}_{a,b}
C^{\A}_{j,k} \,
\m{j}{a} \,  \m{k}{b} }
\prod_{a=1}^{\rG} \prod_{j=1}^{\ell-1}
\Mult{\m{j}{a}+\n{j}{a}}{\m{j}{a}}{q}.
\label{ADEpoly}
\end{equation}
Here the prime signifies the additional constraints
\begin{equation}
\sum_{a=1}^{\rG}
\left(\CG \right)_{r,a}^{-1}  \m{\ell-1}{a}
 \in \Integer,
\label{constronm}
\end{equation}
or, equivalently,
\begin{equation}
\sum_{j=1}^{\ell-1}
\left(C^{\A} \right)_{\ell-1,j}^{-1}  \n{j}{r}
- L
\left(\CG \right)_{r,p}^{-1}
\left(C^{\A} \right)_{\ell-1,1}^{-1}
 \in  \Integer,
\label{constronn}
\end{equation}
with $r=n-1$ for $\cal G$=A$_{n-1}$, $r=n-2$ and $n-1$
for $\cal G$=D$_{n-1}$,
$r=1$ or, equivalently, 5 for E$_6$ and $r=1$ or, equivalently, 7
for E$_7$. Since all entries of
$(C^{\,\mbox{\scriptsize E}_8})^{-1}$
are integers there is no
additional constraint for $\cal G$=E$_8$.

We remark that for $\cal G$=A$_1$ the
fermionic polynomials coincide with a special case of those
defined in refs.~\cite{Melzer,Berkovich}.

\begin{figure}[hbt]
\centerline{\epsffile{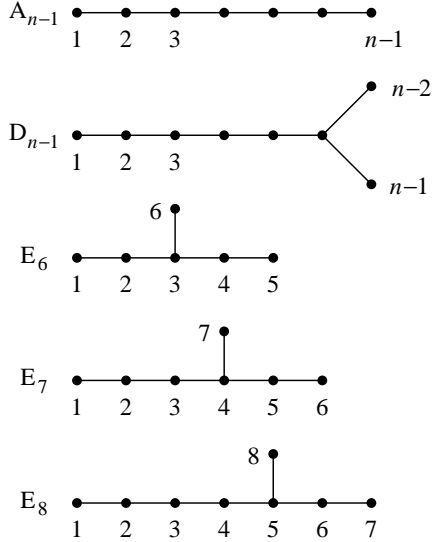}}
\caption{The Dynkin diagrams of the simply-laced Lie algebras
with relevant labelling of the nodes.}
\end{figure}

\nsection{Bosonic A-D-E polynomials}\label{sec3}
As mentioned in the introduction, the bosonic polynomials
all arise naturally in corner transfer matrix calculations
of order parameters of solvable lattice models of RSOS type
\cite{Baxter,ABF}.
One of the key steps in such a calculation is the evaluation
of the {\em one-dimensional configuration sums}
\begin{equation}
X_L(a,b,c) = \sum_{a_2,\ldots,a_L}
q^{\, \sum_{j=1}^L j \, H(a_j,a_{j+1},a_{j+2})} \qquad a_1=a,\: a_{L+1}=b,\:
a_{L+2}=c.
\label{1dcs}
\end{equation}
The function $H$ herein is determined by the Boltzmann weights
of the relevant solvable model, and
each pair $(a_j,a_{j+1})$ is assumed to be admissible in the
sense explained below.

Except for
the case $\cal G$=E$_6$, the bosonic A-D-E polynomials
turn out to coincide with one particular configuration sum.
For E$_6$ it can be defined as a linear combination of
two such sums.

\subsection{A$_{n-1}$}
In this case we have to consider the family of A$_{n-1}^{(1)}$
RSOS models as introduced \cite{JMOa} and solved \cite{JMOb}
by Jimbo, Miwa and Okado (JMO), generalizing the RSOS models
of ABF \cite{ABF} which are recovered for $n=2$.
Before we present JMO's result for the one-dimensional configuration
sums $X_L$ we need some notation.

For the affine Lie algebra A$_{n-1}^{(1)}$ we
let $\Lambda_0,\ldots,\Lambda_{n-1}$ denote the fundamental weights,
$\rho$ the Weyl vector and $\delta$ the null root.
We have the usual inner product on ${\cal H}^{\ast}=
\,\C \Lambda_0 \oplus \ldots \oplus \, \C \Lambda_{n-1}\oplus \, \C \delta$
by
\begin{equation}
\langle \Lambda_{\mu},\Lambda_{\nu} \rangle = \mbox{min}(\mu,\nu)-
\frac{\mu \nu}{n},\qquad
\langle \delta,\delta \rangle =0,
\qquad \langle \Lambda_{\mu},\delta \rangle=1.
\label{innerprod}
\end{equation}
For a general element $a\in {\cal H}^{\ast}$ its level is defined
as $\mbox{lev}(a)=\langle a,\delta \rangle $ and hence, according to
$(\ref{innerprod})$, all fundamental weights have level 1.
A weight $a$ belongs to the set
$P_{+}^{\,\ell}$ of
{\em level-$\ell$ dominant integral weights} if
\begin{equation}
a=\sum_{\mu=0}^{n-1} \Integer_{\geq 0} \, \Lambda_{\mu}
\qquad \mbox{lev}(a)=\ell,
\end{equation}
i.e., $0\leq \langle \rho,a \rangle \leq \ell$.
We finally introduce the vectors $\e_{j}$, $j\in J =\{1,\ldots,n\}$
\begin{equation}
\e_{j} = \Lambda_{j}-\Lambda_{j-1},
\label{evec}
\end{equation}
where we have set $\Lambda_{n}=\Lambda_{0}$.
Clearly $\sum_{j\in J} \e_{j} =0$.
An ordered pair of weights $(a,b)$ both in $P_+^{\,\ell}$
is said to be {\em admissible} and denoted $a\sim b$,
if $b-a=\e_{j}$ for some $j\in J$.

We now come to the definition of the one-dimensional configuration
sums $X_L(a,b,c)$ of the A$_{n-1}^{(1)}$ RSOS models at level $\ell$.
They are given by (\ref{1dcs}), with $a,b,c\in P_{+}^{\,\ell}$ where
the sum is over all sequences $a \sim a_2\sim \ldots
\sim a_{L} \sim b \sim c$ of admissible level-$\ell$
dominant integral weights.
The function $H$ in (\ref{1dcs}) takes the values
\begin{equation}
H(a,a+\e_j,a+\e_j+\e_k)=H(j,k)=
\left\{
\begin{array}{ll}
0 \qquad & 1\leq j <k \leq n \\
& \\
1      & 1\leq k\leq j \leq n .
\end{array} \right.
\end{equation}
Using standard recurrence methods \cite{ABF},
JMO found \cite{JMOb} the following
expression for $X_{L}$:
\begin{equation}
X_L(a,b,b+\e_j) = \sum_{w\in W} \mbox{det}(w) \;
x_L(b+\rho-w(a+\rho),j),
\label{XAn}
\end{equation}
with $W$ the affine Weyl group and, for
$\lambda=\sum_{j\in J} \lambda_j \e_j + z \delta
\in {\cal H}^{\ast}$,
\begin{equation}
x_L(\lambda,j) =
q^{\sum_{k\in J} \lambda_{k}[H(k,j)+(\lambda_k-1)/2]}
\Mult{L}{\lambda}{q}
=q^{|\lambda-\Lambda_{j-1}|^2/2 + (L-j+1)(L-j+1+n)/(2n)}
\Mult{L}{\lambda}{q} .
\end{equation}
The $q$-multinomial coefficient in the definition of $x_L$
reads
\begin{equation}
\Mult{L}{\lambda}{q}=
\left\{\begin{array}{ll}
\displaystyle \frac{(q)_L}{
\prod_{j\in J}(q)_{\lambda_j}} \: ,
\qquad & \displaystyle
\sum_{j\in J} \lambda_j=L, \;
\lambda_j \geq 0\\
& \\
0& \mbox{otherwise.}
\end{array}
\right.
\label{qmulti}
\end{equation}

Though equation (\ref{XAn}) defines bosonic polynomials
for any $a,b$ and $j$, we only consider the special
case $a=b=\ell \Lambda_0$ and $j=1$ and
define\footnote{To explicitly denote the fact that $b$ is
obtained via admissible sequences of $L$ steps on $P_{+}^{\,\ell}$
it may be better to employ the usual Young tableau notation
assigning a tableau of signature $(f_1,\ldots,f_n)$ to each
element of $P_{+}^{\,\ell}$.
In the convention of ref.~\cite{JMOb} we then would have
$a=(0,\ldots,0)$, $b=(L/n,\ldots,L/n)$ and $c=(L/n+1,L/n,\ldots,L/n)$.}
\begin{equation}
B_q^{\An}(L) = q^{-L(L+n)/(2n)} \: X_L(\ell \Lambda_0, \ell \Lambda_0,
\Lambda_1+(\ell-1)\Lambda_0).
\label{BAn}
\end{equation}
Our conjecture is that
$F_q^{\An}(L) =  B_q^{\An}(L)$
for all $L\in \Integer_{\geq 0}$,
$L=0 \ppmod{n}$.
For $n=2$ this was previously conjectured and partially proven
in ref.~\cite{Melzer}. A full proof for $n=2$ can be found in
ref.~\cite{Berkovich}.
To test the conjecture we have generated both
fermionic and bosonic polynomials
$F_q^{\An}(L)$ and $B_q^{\An}(L)$ for all $n+\ell\leq 8$
and for extensive ranges of $L$ using {\em Mathematica}~\cite{Wolfram}.

\subsection{D$_{n-1}$}
For $\cal G$=D$_{n-1}$ we turn to  the hierarchy
of D$_{n-1}^{(1)}$ RSOS models, again introduced by
JMO \cite{JMOa}.  The result for the one-dimensional configuration
sums \cite{DJKMO}
is almost completely analogous to that for A$_{n-1}^{(1)}$
as described in the previous subsection, and hence
we only point out the relevant changes.

First of all, the inner product in the weight space changes to
\begin{eqnarray}
\lefteqn{
\langle \Lambda_{\mu},\Lambda_{\nu} \rangle =\mbox{min}(\mu,\nu),
\qquad
\langle \Lambda_{\mu},\Lambda_{m} \rangle =\case{1}{2}\mu,
\qquad
\langle \Lambda_{m},\Lambda_{m} \rangle = \case{1}{4}(n-1), } \nonumber \\
\lefteqn{
\langle \Lambda_{m},\Lambda_{m'} \rangle = \case{1}{4}(n-3),
\qquad \langle \Lambda_{0,1,n-2,n-1},\delta \rangle=1,
\qquad \langle \Lambda_{2,\ldots,n-3},\delta \rangle=2, }
\end{eqnarray}
with $0\leq \mu \leq n-3$, $m,m'=n-2,n-1$ and $m\neq m'$.
We again introduce vectors $\e_j$, $j\in J=\{\pm1,\ldots,\pm (n-1)\}$,
defined by (\ref{evec}) for $j>0$ and $\e_{-j}=-\e_j$.
Two exceptions to (\ref{evec}) are
$\e_2=\Lambda_2-\Lambda_1-\Lambda_0$
and $\e_{n-2}=\Lambda_{n-2}-\Lambda_{n-3}+\Lambda_{n-1}$,
and we note that $\langle \e_j,\e_k \rangle=jk \: \delta_{|j|,|k|}/|jk|$
for all $j,k\in J$.
A pair $(a,b)$, $a,b\in P_{+}^{\,\ell}$ is admissible
if $b-a \in J$, where $P_{+}^{\,\ell}$
again denotes the level-$\ell$ dominant integral weights.
The values taken by the function $H$ read \cite{DJKMO}
\begin{equation}
H(a,a+\e_j,a+\e_j+\e_k)=H(j,k)=
\left\{
\begin{array}{ll}
0 \qquad &
j<k \mbox{ and } j k >0 \mbox{ or } k<j \mbox{ and } j k <0

\\
& \\
1      & \mbox{otherwise,}
\end{array} \right.
\end{equation}
together with the exceptions
$H(1,-1)=-1$ and $H(-(n-1),n-1)=0$.

Again by application of recurrence methods it was
shown in ref.~\cite{DJKMO}
that $X_L$ can be expressed as the sum over the
affine Weyl group as defined in (\ref{XAn}),
with $x_L$ therein given by
\begin{equation}
x_L(\lambda,j) =
\sum_{\eta_j-\eta_{-j}=\lambda_j}
q^{-\eta_p\eta_{-p}+\sum_{k\in J} \eta_{k}[H(k,j)+(\eta_k-1)/2]}
\Mult{L}{\eta}{q},
\end{equation}
with $p=1$ for $j<0$ and $p=n-1$ for $j>0$, and with the
same definition of the multinomial coefficient as in (\ref{qmulti}).

As in the previous case, we only consider the bosonic polynomials
obtained by specializing
$a=b=\ell \Lambda_0$ and $j=1$ in (\ref{XAn}), and set
\begin{equation}
B_q^{\Dn}(L) = q^{-L/2} \: X_L(\ell \Lambda_0, \ell \Lambda_0,
\Lambda_1+(\ell-1)\Lambda_0).
\label{BDn}
\end{equation}
We then conjecture that
$F_q^{\Dn}(L) =  B_q^{\Dn}(L)$
for all even $L\in \Integer_{\geq 0}$.
This conjecture has again been tested extensively for all
$n+\ell\leq 8$
using {\em Mathematica}~\cite{Wolfram}.

\subsection{E$_{6,7,8}$}
We now come to the exceptional simply-laced Lie algebras:
E$_6$, E$_7$ and E$_8$.
Unfortunately, in this case we do not have an algebraic
formulation of the bosonic polynomials, and as an immediate
consequence we only have conjectures for $\ell=2$.

In all three cases to follow we use the following definition of the
$q$-multinomial
\begin{equation}
\Mult{N}{m_1,m_2}{q}=\left\{
\begin{array}{ll}
\displaystyle \frac{(q)_N}{(q)_{m_1}
(q)_{m_2}(q)_{N-m_1-m_2}} \qquad & 0\leq m_1+m_2 \leq N,
\; m_1,m_2\geq 0 \\
& \\
0& \mbox{otherwise.}
\end{array}
\right.
\end{equation}

\subsubsection{E$_6$}
Define the polynomials
\begin{eqar}
\lefteqn{
B^{\, \mbox{\scriptsize E}_6}_q(L) =
\sum_{j,k=-\infty}^{\infty} \left\{
q^{42j^2+j+k(k+14j)}\Mult{L}{k,k+14j}{q}-
q^{42j^2+13j+1+k(k+14j+2)}\Mult{L}{k,k+14j+2}{q} \right\} }
\nonumber \\
& & \label{BE6}\\
\lefteqn{ \quad + q^5  \sum_{j,k=-\infty}^{\infty} \left\{
q^{42j^2+29j+k(k+14j+5)} \Mult{L}{k,k+14j+5}{q}-
q^{42j^2+41j+5+k(k+14j+7)}\Mult{L}{k,k+14j+7}{q} \right\}. }
\nonumber
\end{eqar}
It is to be noted that 5 is the integer value of the
weight $\Delta_{5,1}^{(6,7)}$ and hence that the above form
corresponds to a finitization of the {\em extended}
identity character (with respect to the Virasoro algebra).
The bosonic expression (\ref{BE6}) can be identified with the
normalized sum of the two configurations sums ($X_L(1,1,2)$ and $X_L(1,6,5)$
in the notation of ref.~\cite{WPSN}) of the dilute A$_6$ lattice
model \cite{unpublished}.

Our conjecture is that
$F^{\, \mbox{\scriptsize E}_6}_q(L) =
B^{\, \mbox{\scriptsize E}_6}_q(L)$,
for all
$L\in \Integer_{\geq 0}$.
We note that for $\ell=2$ the restriction
(\ref{constronm}) is implied by
the constraint system (\ref{TBAconstr}) and as
a result we can drop
the prime in the sum (\ref{ADEpoly}).
We have checked the correctness of the E$_6$
polynomial identity for all $L\leq 33$ by direct expansion.

\subsubsection{E$_7$}
Define the polynomials
\begin{eq}
B^{\, \mbox{\scriptsize E}_7}_q(L) =
\sum_{j,k=-\infty}^{\infty} \left\{
q^{20j^2+j+k(k+10j)}\Mult{L}{k,k+10j}{q}-
q^{20j^2+9j+1+k(k+10j+2)}\Mult{L}{k,k+10j+2}{q} \right\}.
\end{eq}
This bosonic expression is related to one of the
configuration sums ($X_L(1,1,2)$ in the notation of \cite{WPSN})
of the dilute A$_4$ model \cite{unpublished}.

Our conjecture is that
$F^{\, \mbox{\scriptsize E}_7}_q(L) =
B^{\, \mbox{\scriptsize E}_7}_q(L)$,
for all
$L\in \Integer_{\geq 0}$.
Tests have again confirmed the
polynomial identity for all $L\leq 37$.

\subsubsection{E$_8$}
Define the polynomials
\begin{equation}
B^{\, \mbox{\scriptsize E}_8}_q(L) =
\sum_{j,k=-\infty}^{\infty} \left\{
q^{12j^2+j+k(k+8j)}\Mult{L}{k,k+8j}{q}-
q^{12j^2+7j+1+k(k+8j+2)}\Mult{L}{k,k+8j+2}{q} \right\}.
\end{equation}
Comparison with the one-dimensional configuration sums
for the dilute A$_3$ model in regime 2 as computed in
ref.~\cite{WPSN} shows that $
B^{\, \mbox{\scriptsize E}_8}_q(L) = X_L(1,1,2)$.

The assertion is now that
$F^{\, \mbox{\scriptsize E}_8}_q(L) =
B^{\, \mbox{\scriptsize E}_8}_q(L)$,
for all
$L\in \Integer_{\geq 0}$.
The E$_8$ polynomial identity has in fact been
proven \cite{WP}, and is directly related to the
E$_8$ structure of the critical Ising model in
a magnetic field \cite{Zamolodchikov,BNW}.

\nsection{A-D-E Rogers--Ramanujan identities}\label{sec4}
In this section we present the Rogers--Ramanujan type identities
that follow by taking the $L\to\infty$ limit in the
various polynomial identities listed in the previous section.
To put the results in the context of coset conformal field
theories based on the GKO construction of equation (\ref{GKO}),
we first give a brief reminder on theta functions and
{\em branching rules},
see e.g., refs.~\cite{Kac,KP,GKO,JMOb,BBSS,CR}.

\subsection{Branching functions}
For $\mu,z\in \overline{\cal H}^{\ast} $
the classical $\cal G$ theta function of
characteristic $\mu$ and degree $m$
is defined as\footnote{As in the rest of the paper
$\cal G$ denotes a simply-laced Lie algebra.}
\begin{equation}
\Theta_{\mu,m}(z,\tau) =
\sum_{\alpha \, \in\,  Q + \frac{\mu}{m}}
\e^{m \pi \is \tau |\alpha|^2
-2\pi \is m \langle \alpha,z \rangle},
\label{Gtheta}
\end{equation}
with $Q$ the root lattice of $\cal G$.
Then,
according to the Weyl-Kac formula \cite{KP},
the character of highest weight representation $a\in P_{+}^{\,\ell}$ of
a ${\cal G}^{(1)}$ Kac-Moody algebra
is given by
\begin{equation}
\chi_{a,\ell}(z,\tau)  =
\frac{\displaystyle \sum_{w\in \overline W} \mbox{sgn}(w) \;
\Theta_{w(\overline{a}+\overline{\rho}),g+\ell}(z,\tau)}
{\displaystyle \sum_{w\in \overline W} \mbox{sgn}(w) \;
\Theta_{w(\overline{\rho}),g}(z,\tau)} \: ,
\end{equation}
with, $\overline W$ the Weyl group and  $g$ the dual
Coxeter number of $\cal G$.

We now come to the definition of the {\em branching functions}
$b_{a,b}^{\, (t)}(\tau)$ for dominant integral
weights $a\in P_{+}^{\,\ell-1}$,
$b\in P_{+}^{\,\ell}$ and $t\in P_{+}^{1}$,
$\overline{t}=\overline{b}-\overline{a}
\ppmod{Q}$, via the decomposition or branching rule
\begin{equation}
\chi_{a,\ell-1}(z,\tau) \:
\chi_{t,1}(z,\tau) = \sum_{b\in P_{+}^{\,\ell}}
b_{a,b}^{\, (t)}(\tau) \:
\chi_{b,\ell}(z,\tau).
\label{brule}
\end{equation}
The branching functions can be expressed in terms of
the $\cal G$ theta functions as
\begin{equation}
b_{a,b}^{\, (t)}(\tau) =
\frac{1}{[\eta(q)]^{\,\rG}}
\sum_{w \in \overline{W}} \mbox{sgn}(w) \;
\Theta_{- (p+1)(\overline{a}+\overline{\rho})
+ p w(\overline{b}+\overline{\rho}),p(p+1)}(0,\tau),
\end{equation}
with $p=\ell+g-1$.
{}From the definition (\ref{Gtheta}) of the theta functions
this can be rewritten in a form generalizing equation (\ref{Xrs})
for the unitary minimal Virasoro characters \cite{CR}
\begin{equation}
b_{r,s}^{\, (t)}(q) =
\frac{1}{[\eta(q)]^{\,\rG}}
\sum_{\alpha \in Q} \, \sum_{w\in \overline{W}} \mbox{sgn}(w) \; \;
q^{\displaystyle
\case{1}{2} \, p (p+1) \left| \alpha -\frac{(p+1)\overline{r}
- p \, w(\overline{s})}{p(p+1)}\right|^2},
\label{bf}
\end{equation}
with $q=\exp(2\pi \ib \tau)$, $r=a+\rho$ and $s=b+\rho$.
The lowest order term in this expression occurs for $w$=id, $\alpha=0$,
and thus
\begin{equation}
b_{r,s}^{(t)}(q) = q^{\Delta^{(p,p+1)}_{r,s}-c/24}
\sum_{n=0}^{\infty} a_n q^n \qquad \quad a_n\in \Integer_{\geq 0}\, ,
\label{bfexp}
\end{equation}
generalizing (\ref{cc}) and (\ref{weights}) to
\begin{eqnarray}
c&=& \mbox{rank }{\cal G} \left(1-\frac{g(g+1)}{p(p+1)}
\right) \label{ccG} \\
& & \nonumber \\
\Delta_{r,s}^{(p,p+1)} &=&
\frac{\displaystyle [(p+1)r-ps]^2-\frac{g \: \mbox{dim }{\cal G}}{12}}
{2p(p+1)}\; .
\label{delG}
\end{eqnarray}

\subsection{Rogers--Ramanujan type identities}
We now consider the $L\to\infty$ limit of the
polynomial identities as conjectured in sections~\ref{sec2} and
\ref{sec3}.

Eliminating the $\n{j}{a}$ from the fermionic polynomials
(\ref{ADEpoly}) using (\ref{TBAconstr}), and then letting
$L\to \infty$ by $\lim_{L\to\infty} \Mults{N}{m}{q} = 1/(q)_m$, yields
\begin{eq}
\lim_{L\to\infty}F_q^{\,\cal G}(L) \equiv F_q^{\,\cal G}=
\left. \sum \right.^{'}q^{\; \displaystyle \case{1}{2}
\sum_{a,b=1}^{\rG} \sum_{j,k=1}^{\ell-1}
\left(\CG\right)^{-1}_{a,b}
C^{\A}_{j,k} \,
\m{j}{a} \,  \m{k}{b} }
\prod_{a=1}^{\rG} \frac{1}{(q)_{\m{1}{a}}} \; \prod_{j=2}^{\ell-1}
\Mult{\displaystyle \m{j}{a}
+ P_j^{(a)} }{\m{j}{a}}{q} \! \! .
\label{finf}
\end{eq}
Here we have introduced the symbol $P_j^{(a)}$ to mean
\begin{equation}
P_j^{(a)} = \left\{
\begin{array}{ll}
\infty & j=1 \\ & \\
\displaystyle -\sum_{b=1}^{\rG} \sum_{k=1}^{\ell-1}
\left(\CG\right)_{a,b}^{-1}
C^{\A}_{j,k}
\m{k}{b} \qquad  & j=2,\ldots,\ell-1,
\end{array} \right.
\label{defP}
\end{equation}
with $P_1^{(a)}$ included for later convenience.
Some caution has to be taken in interpreting the primed sum in
(\ref{finf}). No longer is the sum over all solutions to
the constraint system (\ref{TBAconstr}), but simply over
all $\m{j}{a}\in \Integer_{\geq 0}$. The prime still
denotes the condition (\ref{constronm}) where it is to be noted that
we can no longer drop the prime for $\cal G$=E$_6$.

The $L\to\infty$ limit of the bosonic polynomials $B_q^{\An}(L)$
defined in (\ref{BAn}) was
considered in ref.~\cite{JMOb}.
The result reads
\begin{equation}
\lim_{L\to \infty} B_q^{\An}(L) \equiv B_q^{\An} =
q^{c/24} \; b_{r,s}^{(t)}(q)
\qquad \quad \left\{
\begin{array}{l}
r=(\ell-1)\Lambda_0+\rho \\
s=\ell \Lambda_0+\rho \\
t=\Lambda_0,
\end{array}
\right.
\label{Binf}
\end{equation}
where the branching function is that of the previous subsection
based on $\cal G$=A$_{n-1}$.

In taking the infinite limit of the D$_{n-1}$ polynomials
(\ref{BDn}) we get exactly the same result as in
(\ref{Binf}) with A$_{n-1}$ replaced by D$_{n-1}$ \cite{DJKMO}
and interpreting the branching function as that of
$\cal G$=D$_{n-1}$.

For the exceptional cases we only have $\ell=2$
and a simplification occurs as the exceptional
branching functions at level 2
reproduce the ordinary Virasoro characters $\chi_{r,s}^{(p,p+1)}$
in the bosonic representation of equation (\ref{Xrs}).
To illustrate this consider for example $\cal G$=E$_8$.
Then $P_{+}^{1}=\{\Lambda_0\}$ and
$P_{+}^{2}=\{2\Lambda_0,\Lambda_1,\Lambda_7\}$ and we
compute from (\ref{ccG}) $c=1/2$ and from (\ref{delG})
\begin{equation}
\begin{array}{rcl}
\Delta_{\Lambda_0+\rho,2\Lambda_0+\rho}^{(31,32)}=& \! \! \! 0 \! \! \! &=
\Delta_{1,1}^{(3,4)} \\ & & \\
\Delta_{\Lambda_0+\rho,\Lambda_1+\rho}^{(31,32)}=&
\displaystyle \! \! \! \frac{1}{16} \! \! \! &=
\Delta_{1,2}^{(3,4)} \\ & & \\
\Delta_{\Lambda_0+\rho,\Lambda_7+\rho}^{(31,32)}=&
\displaystyle \! \! \! \frac{1}{2} \! \! \! &=
\Delta_{2,1}^{(3,4)}.
\end{array}
\end{equation}
Here the weights on the left-hand-side are to be understood as
those of (\ref{delG}) based on $\cal G$=E$_8$ and
the weights on the right-hand-side as
those of equation (\ref{weights}) or, equivalently, as those of (\ref{delG})
with $\cal G$=A$_1$.
Hence writing the three infinite forms of the bosonic
polynomials in terms of Virasoro characters
yields \cite{WPSN,unpublished}
\begin{equation}
\lim_{L\to\infty}
B^{\, \mbox{\scriptsize E}_n}_q(L) \equiv
B^{\, \mbox{\scriptsize E}_n}_q =
\left\{
\begin{array}{ll}
q^{c/24} \left( \chi_{1,1}^{(6,7)} + \chi_{5,1}^{(6,7)} \right)
\qquad & n=6 \\
q^{c/24} \chi_{1,1}^{(4,5)} & n=7 \\
q^{c/24} \chi_{1,1}^{(3,4)} & n=8 .
\end{array}
\right.
\label{delisdel}
\end{equation}

With the above definitions (\ref{finf})-(\ref{delisdel})
the $\cal G$ Rogers--Ramanujan type identities
can be written as
\begin{equation}
F_q^{\, \cal G} = B_q^{\, \cal G}.
\label{GRR}
\end{equation}
For $\ell=2$ and arbitrary $\cal G$ this has been conjectured
by Kedem {\em et al.} in ref.~\cite{KKMMa}.
For $\cal G$=A$_1$ and  arbitrary $\ell$
this was again conjectured by Kedem {\em et al.}, this time in
ref.~\cite{KKMMb}.
Also the general form of (\ref{GRR}) can be inferred from
refs.~\cite{KKMMa,KKMMb}, but details such as the precise
form of the restrictions (\ref{constronm}) on
the sum in (\ref{finf}) were not
conjectured in full generality.

\nsection{Level-rank duality}\label{sec5}
An important consideration in the confirmation of the
conjectured polynomial identities is
their behaviour under the transformation
$q \to 1/q$.
In particular, for the $\cal G$=A$_{n-1}$ case
we show that the identity
$F^{\,\cal G}_q(L) =
B^{\,\cal G}_q(L)$
displays
the correct invariance under the
simultaneous transformations
\begin{equation}
\ell  \leftrightarrow n, \qquad
q  \leftrightarrow  1/q \qquad
\label{levelrank}
\end{equation}
provided we choose $
L \left(C^{\mbox{\tiny A}_{n-1}} \right)_{n-1,1}^{-1}
\left(C^{\mbox{\tiny A}_{\ell-1}} \right)_{\ell-1,1}^{-1} = L/n\ell
\in \Integer $.

\subsection{The transformation $q\to 1/q$}
Let us first consider how the general fermionic sum (\ref{ADEpoly})
transforms under inversion of $q$.
To do so we rewrite (\ref{TBAconstr}) in a form expressing
the number of anti-particles in terms of the number
of particles, and vice versa
\begin{eqnarray}
\n{j}{a} &=& L \: \delta_{j,1}
\left(\CG\right)_{a,p}^{-1} -
\sum_{b=1}^{\rG} \sum_{k=1}^{\ell-1}
\left(\CG \right)_{a,b}^{-1}
C^{\A}_{j,k} \,
\m{k}{b} \label{nasm}  \\
\m{j}{a} &=& L \: \delta_{a,p}
\left(C^{\A} \right)_{j,1}^{-1} -
\sum_{b=1}^{\rG} \sum_{k=1}^{\ell-1}
\CG_{a,b}
\left(C^{\A} \right)_{j,k}^{-1} \,
\n{k}{b} . \label{masn}
\end{eqnarray}
Now, upon using the inversion
\begin{equation}
\Mult{N}{m}{1/q} = q^{m(m-N)} \Mult{N}{m}{q},
\end{equation}
we find that the exponent of $q$ in the
expression for $F_{1/q}^{\,\cal G}(L)$ reads
\begin{equation}
-\frac{1}{2}
\sum_{a,b=1}^{\rG} \sum_{j,k=1}^{\ell-1}
\left(\CG\right)^{-1}_{a,b}
C^{\A}_{j,k}
\m{j}{a} \, \m{k}{b} -
\sum_{a=1}^{\rG} \sum_{j=1}^{\ell-1}
\m{j}{a} \, \n{j}{a}.
\end{equation}
Substituting (\ref{nasm}) to eliminate
$\n{j}{a}$ this becomes
\begin{eqnarray}
\lefteqn{
-\frac{L^2}{2}
\left(\CG\right)^{-1}_{p,p}
\left(C^{\A}\right)^{-1}_{1,1}} \nonumber \\
& & \nonumber \\
& & + \frac{1}{2}
\sum_{a,b=1}^{\rG} \sum_{j,k=1}^{\ell-1}
\left(\CG\right)^{-1}_{a,b}
C^{\A}_{j,k}
\left[\m{j}{a}-L\: \delta_{a,p}
\left(C^{\A}\right)^{-1}_{j,1} \right]
\left[\m{k}{b}-L\: \delta_{b,p}
\left(C^{\A}\right)^{-1}_{k,1} \right].
\end{eqnarray}
Then, using (\ref{masn}), we arrive at
\begin{equation}
-\frac{L^2}{2}
\left(\CG\right)^{-1}_{p,p}
\left(C^{\A}\right)^{-1}_{1,1}+
\frac{1}{2}
\sum_{a,b=1}^{\rG} \sum_{j,k=1}^{\ell-1}
\CG_{a,b}
\left(C^{\A}\right)^{-1}_{j,k}
\n{j}{a}  \, \n{k}{b}.
\end{equation}
We now carry out a transformation of variables.
Replacing $\m{j}{a}\to \n{a}{j}$ and
$\n{j}{a}\to \m{a}{j}$ followed by $j\leftrightarrow a$ and
$k \leftrightarrow b$ yields
\begin{eqnarray}
\lefteqn{
F_{1/q}^{\,\cal G}(L) = q^{\displaystyle -\case{L^2}{2}
\left(\CG\right)^{-1}_{p,p}
\left(C^{\A}\right)^{-1}_{1,1} } } \nonumber \\
& & \nonumber \\
& & \qquad
\times \left. \sum \right.^{'}q^{\; \displaystyle \case{1}{2}
\sum_{a,b=1}^{\ell-1} \sum_{j,k=1}^{\rG}
\left(C^{\A}\right)^{-1}_{a,b}
\CG_{j,k} \,
\m{j}{a} \,  \m{k}{b} }
\prod_{a=1}^{\ell-1} \prod_{j=1}^{\rG}
\Mult{\m{j}{a}+\n{j}{a}}{\m{j}{a}}{q}.\label{ADEdual}
\end{eqnarray}
Of course, now the sum is over all solutions of
\begin{equation}
\m{j}{a} + \n{j}{a} =
\frac{1}{2} \left[ L \: \delta_{a,1} \delta_{j,p} +
\sum_{b=1}^{\ell-1}
{\cal I}^{\A}_{a,b}
\n{j}{b} +
\sum_{k=1}^{\rG} {\cal I}^{\, \cal G}_{j,k} \:
\m{k}{a} \right],
\label{TBAdual}
\end{equation}
where the prime denotes the additional condition
\begin{equation}
\sum_{a=1}^{\ell-1}
\left(C^{\A} \right)_{\ell-1,a}^{-1}  \m{r}{a}
- L
\left(C^{\A} \right)_{\ell-1,1}^{-1}
\left(\CG \right)_{r,p}^{-1}
 \in  \Integer.
\label{constronndual}
\end{equation}
or, equivalently,
\begin{equation}
\sum_{j=1}^{\rG}
\left(\CG \right)_{r,j}^{-1}  \n{j}{\ell-1}
 \in \Integer.
\label{constronmdual}
\end{equation}
In the following we denote the fermionic sum
(\ref{ADEdual}) without the irrelevant factor
in front of the summation symbol by
$G_q^{\,\cal G}(L)$.
We note the apparent similarity of $F_q^{\, \cal G}(L)$
and $G_q^{\, \cal G}(L)$
under interchange
of $\cal G$ and A$_{\ell-1}$ in either one of the two expressions.

\subsection{The case $\cal G$=A$_{n-1}$: level-rank duality}
We now proceed to consider the inversion properties of the
polynomial expressions in the case $\cal G$=A$_{n-1}$ only.

To explicitly exhibit the dependence of the fermionic polynomials
on A$_{n-1}$ as well as on A$_{\ell-1}$, we write
$F_q^{\An}(L) = F_q^{(n,\ell)}(L)$.
Then, since $p=1$ and $r=n-1=\mbox{rank}$ A$_{n-1}$,
we clearly have that $F_q^{(n,\ell)}(L)=G_q^{(\ell,n)}(L)$,
given we choose $L$ such that $L=0 \ppmod{n \ell}$.
This duality of the fermionic polynomials under the transformation
(\ref{levelrank}) is an example of so-called {\em level-rank}
duality \cite{KN}.
Of course, for our conjecture
$F_q^{(n,\ell)}(L) =  B_q^{(n,\ell)}(L)\; (=B_q^{\An}(L))$
to be correct, also the bosonic polynomials $B_q^{\An}(L)$ in
(\ref{BAn})
must remain invariant under transformation (\ref{levelrank}).
As pointed out in ref.~\cite{JMOb} this is indeed the case.
Apart from the overall factor
\begin{equation}
q^{\displaystyle -\case{1}{2}L^2
\left(\CG\right)^{-1}_{1,1}
\left(C^{\A}\right)^{-1}_{1,1} } =
q^{\displaystyle -\frac{L^2(n-1)(\ell-1)}{2n\ell}} \quad ,
\end{equation}
we again have level-rank duality
provided $L=0 \ppmod{n \ell}$.

It is in fact precisely this level-rank duality
for $\cal G$=A$_{n-1}$
that
led us to the conjecture (\ref{ADEpoly}).
That is, from the already known and proven \cite{Melzer,Berkovich}
polynomial identity for A$_1$ at level $\ell-1$ one can
immediately infer a polynomial identity for
A$_{\ell-1}$ at level 2. To then write down
polynomial identities for general rank, level and $\cal G$ is a
matter of straightforward generalization.

\nsection{Dilogarithm identities}\label{sec6}
It was first shown by Richmond and Szekeres~\cite{RS}
that in studying
Rogers--Ramanujan identities in the limit
$q\to 1$, one obtains identities involving
dilogarithms.
We here show that all A-D-E identities  of the
previous section indeed
yield the expected dilogarithm identities.
In particular, we will show that for $q\to 1^{-}$
the Rogers--Ramanujan identities imply the identities
\begin{equation}
s^{\, \cal G}(\ell-1)
+ s^{\, \cal G}(1)
- s^{\, \cal G}(\ell) = c ,
\label{ADEdilog}
\end{equation}
corresponding to the coset conformal field theories
(\ref{GKO}) with central charges $c$ as given by
(\ref{ccG}).
Here $s^{\, \cal G}$ is defined as a sum
over the Rogers dilogarithm function \cite{Lewin}
\begin{equation}
L(z) = \frac{1}{2} \int_0^z \left[\frac{\log(1-\zeta)}{\zeta}
+ \frac{\log\zeta}{1-\zeta} \right] \d{\zeta}
= -\int_0^z \frac{\log(1-\zeta)}{\zeta}\: \d{\zeta}
+\frac{1}{2} \log z \, \log (1-z)
\label{dilog}
\end{equation}
as follows:
\begin{equation}
s^{\,\cal G}(\ell) = \frac{6}{\pi^2}
\sum_{a=1}^{\rG} \sum_{j=1}^{\ell}
L\left(\xi_j^{(a)}\right).
\label{defofsG}
\end{equation}
The numbers $\xi_{\ell}^{(a)}$ in the above sum
are the solutions to the TBA equations \cite{BR,KM,AlB,Ravanini}
\begin{equation}
\sum_{b=1}^{\rG} \CG_{a,b}\; \log \left(1-\xi_j^{(b)}\right) =
\sum_{k=1}^{\ell-1} C^{\A}_{j,k} \, \log \left(\xi_k^{(a)} \right),
\label{TBAxi}
\end{equation}
where by definition $\xi_{\ell}^{(a)}=1$.

Before proceeding to derive (\ref{ADEdilog}) from the
Rogers--Ramanujan identities (\ref{GRR}),
let us mention that  the
above dilogarithm identities
were first conjectured in ref.~\cite{BR}
in the computation of  central charges of A-D-E TBA systems.
For the occurrence of these same identities in related work
on TBA, see e.g., refs.~\cite{KM,AlB,Ravanini}.
A proof of (\ref{ADEdilog}) for $\cal G$=A$_{n-1}$ has
been given by Kirillov in ref.~\cite{Kirillov}.

To establish (\ref{ADEdilog}), we follow
the working of Nahm {\em et al.}~\cite{NRT}
(see also refs.~\cite{RS,DKKMM,Kirillovb})
in evaluating $F_q^{\, \cal G}$, $q\to 1^{-}$
using steepest descent.
Writing
\begin{equation}
F_q^{\, \cal G} = \sum_{\{\m{j}{a}\}} f_q^{\, \cal G}(\{\m{j}{a}\})
= \sum_{M=0}^{\infty} a_M \, q^M
\end{equation}
we have
\begin{equation}
a_{M-1} = \frac{1}{2\pi \ib} \sum_{\{m_j^{(a)}\}} \oint q^{-M}
f_q^{\, \cal G}(\{\m{j}{a}\}) \, \d{q}.
\label{am}
\end{equation}
Treating the $\m{j}{a}$ as continuous variables,
we now approximate the integration kernel by
\begin{eqnarray}
\log \left( q^{-M}
f_q^{\, \cal G}(\{\m{j}{a}\}) \right) &\approx &
\left( \frac{1}{2}
\sum_{a,b=1}^{\rG} \sum_{j,k=1}^{\ell-1}
\left(\CG\right)^{-1}_{a,b}
C^{\A}_{j,k}
\m{j}{a} \, \m{k}{b} - M \right) \log q \nonumber \\
& & +\sum_{a=1}^{\rG} \sum_{j=1}^{\ell-1}
\left(
\int_0^{P_j^{(a)}+\m{j}{a}} -
\int_0^{P_j^{(a)}} -
\int_0^{\m{j}{a}} \right)
\log \left( 1-q^t \right) \d{t}.
\label{fq}
\end{eqnarray}
Differentiating with respect to $m_j^{(a)}$ to find the
saddle point results in the
TBA equations (\ref{TBAxi}),
with $\xi_j^{(a)}$ defined by
\begin{equation}
\xi_j^{(a)} = \frac{q^{\m{j}{a}} \left( 1-q^{P_j^{(a)}}\right)}
{1-q^{P_j^{(a)}+\m{j}{a}}}.
\label{defxi}
\end{equation}
Using the definition (\ref{dilog}) of the dilogarithm function $L$,
the simple relation
\begin{equation}
L(z)+L(1-z)=L(1)=\frac{\pi^2}{6}
\end{equation}
and the pentagonal
relation~\cite{Lewin}
\begin{equation}
L(1-x)+L(1-y)-L(1-xy) = L\left(\frac{x(1-y)}{1-xy}\right)
- L\left(\frac{1-y}{1-xy}\right),
\end{equation}
yields
\begin{eqnarray}
\lefteqn{
\log \left( q^{-M}
f_q^{\, \cal G}(\{\m{j}{a}\}) \right) } \nonumber \\
&  &\approx  -M \log q + \frac{1}{\log q} \:
\sum_{a=1}^{\rG} \sum_{j=1}^{\ell-1}
\left[
L\left(\xi_j^{(a)}\right) - L\left(\eta_j^{(a)}\right) \right]
\nonumber \\
& & \quad +
\frac{1}{2}
\sum_{a=1}^{\rG} \sum_{j=1}^{\ell-1} \left\{
\log q \sum_{b=1}^{\rG} \sum_{k=1}^{\ell-1}
\left(\CG\right)^{-1}_{a,b}
C^{\A}_{j,k}
\m{j}{a} \, \m{k}{b} \right.
\label{beforecancel} \\
& &\quad
\left.
\vphantom{\left(\CG\right)^{-1}_{a,b}}
+ \left(P_j^{(a)} + \m{j}{a} \right) \log
\left( 1-q^{P_j^{(a)} + \m{j}{a}} \right)
- P_j^{(a)} \log \left(1-q^{P_j^{(a)}} \right)
-\m{j}{a} \log \left(1-q^{\m{j}{a}} \right)  \right\}. \nonumber
\end{eqnarray}
The variables $\eta_j^{(a)}$ in this expression are given by
\begin{equation}
\eta_j^{(a)} = \frac{1-q^{P_j^{(a)}}}
{1-q^{P_j^{(a)}+\m{j}{a}}}
\end{equation}
and, by (\ref{defP}) and (\ref{defxi}), they satisfy
\begin{eqnarray}
\lefteqn{\eta_1^{(a)}  =  1} \nonumber
\nonumber \\ & &  \nonumber \\
\lefteqn{
\sum_{b=1}^{\rG} \CG_{a,b}\; \log \left(1-\eta_{j+1}^{(a)}\right) =
\sum_{k=1}^{\ell-2} C^{\, \mbox{\scriptsize A}_{\ell-2}}_{j,k} \,
\log \left(\eta_{k+1}^{(b)} \right) \qquad j=1,\ldots,\ell-2. }
\label{TBAeta}
\end{eqnarray}
Clearly, the $\eta_j^{(a)}$, $j=2,\ldots,\ell-1$ satisfy the
same TBA equations as the $\xi_{j}^{(a)}$, $j=1,\ldots,\ell-1$,
but with $\ell$ replaced by $\ell-1$.

The following elementary manipulations serve to
show that the last two lines in (\ref{beforecancel}) cancel
as a consequence of the TBA equations (\ref{TBAxi}):
\begin{eqar}
\lefteqn{\log q \sum_{a,b=1}^{\rG} \sum_{j,k=1}^{\ell-1}
\left(\CG\right)^{-1}_{a,b} C^{\A}_{j,k}
\m{j}{a} \, \m{k}{b} } \nonumber \\
& &  =
\sum_{a,b=1}^{\rG} \sum_{j,k=1}^{\ell-1}
\left(\CG\right)^{-1}_{a,b} C^{\A}_{j,k} \m{j}{a}
\log \left( \frac{\xi_k^{(b)}}{\eta_k^{(b)}} \right) \nonumber \\
& & =
\sum_{a=1}^{\rG} \sum_{j=1}^{\ell-1}
\m{j}{a}
\log \left( 1-\xi_j^{(a)} \right) -
\sum_{a,b=1}^{\rG} \sum_{j,k=1}^{\ell-1}
\left(\CG\right)^{-1}_{a,b} C^{\A}_{j,k} \m{j}{a}
\log \left( \eta_k^{(b)} \right) \nonumber \\
& & =
\sum_{a=1}^{\rG} \sum_{j=1}^{\ell-1} \left\{
P_j^{(a)} \log \left(1-q^{P_j^{(a)}} \right)
+ \m{j}{a} \log \left(1-q^{\m{j}{a}} \right)
- \left(P_j^{(a)} + \m{j}{a} \right) \log
\left( 1-q^{P_j^{(a)} + \m{j}{a}} \right)\right.  \nonumber \\
& & \qquad \qquad   \qquad
\left. - \left( P_j^{(a)} +
\sum_{b=1}^{\rG} \sum_{k=1}^{\ell-1}
\left(\CG\right)^{-1}_{a,b} C^{\A}_{j,k} \m{k}{b} \right)
\log\left( \eta_j^{(a)} \right)  \right\} \\
& & =
\sum_{a=1}^{\rG} \sum_{j=1}^{\ell-1} \left\{
P_j^{(a)} \log \left(1-q^{P_j^{(a)}} \right)
+\m{j}{a} \log \left(1-q^{\m{j}{a}} \right)
-\left(P_j^{(a)} + \m{j}{a} \right) \log
\left( 1-q^{P_j^{(a)} + \m{j}{a}} \right) \right\}. \nonumber
\end{eqar}
As a result of this we have
\begin{eqnarray}
\log \left( q^{-M}
f_q^{\, \cal G}(\{\m{j}{a}\}) \right) &\approx &
-M \log q + \frac{1}{\log q} \:
\sum_{a=1}^{\rG} \sum_{j=1}^{\ell-1}
\left[
L\left(\xi_j^{(a)}\right) - L\left(\eta_j^{(a)}\right) \right]
\nonumber \\
&=&
-M \log q - \frac{\pi^2}{6 \log q}
\left[
s^{\, \cal G}(\ell-1)
+ s^{\, \cal G}(1)
- s^{\, \cal G}(\ell)  \right].
\label{simple}
\end{eqnarray}
Finally we have to fix the value of $q$ at the saddle point.
{}From $\frac{\mbox{\small d}}{\mbox{\small d}\textstyle q}
\log f_q=0$ this is found to be
\begin{equation}
(\log q)^2 = \frac{\pi^2}{6 M}
\left[
s^{\, \cal G}(\ell-1)
+ s^{\, \cal G}(1)
- s^{\, \cal G}(\ell)  \right].
\end{equation}
Returning to the expression (\ref{am}) hence yields the following
result for the asymptotics of $a_M$
\begin{equation}
a_M \sim \exp\left(2 \pi \sqrt{
\left[
s^{\, \cal G}(\ell-1)
+ s^{\, \cal G}(1)
- s^{\, \cal G}(\ell)  \right]
M/6}\right).
\label{sss}
\end{equation}

To actually obtain the identity (\ref{ADEdilog})
we have to show that in addition to (\ref{sss}) we
also have $a_M\sim \exp(2 \pi \sqrt{cM/6})$
with $c$ given by (\ref{ccG}).
However, from the automorphic properties of the $\cal G$ theta functions
this can indeed be established \cite{KP,DV,JMOb,CR}.
That is, by carrying out the modular transformation
$\tau \to -1/\tau$ which relates
the $q\to 1$ to the $q\to 0$ limit of the branching functions
(\ref{bf}) then using for $(q)_{\infty}^{-1}=\sum a_M q^M$
that $\log a_M \sim 2\pi \sqrt{M/6}$,
the central charge follows as given in (\ref{ccG}).
In fact, it is precisely the prefactor $q^{\Delta_{r,s}^{(p,p+1)}-c/24}$
in the
expansion (\ref{bfexp}) of the branching functions that ensures proper
modular invariance of the branching rule (\ref{brule}).

Before ending this section let us make some concluding remarks.
We have presented the dilogarithm identities (\ref{ADEdilog})
as special sums over the solution to the TBA equations
(\ref{TBAxi}) avoiding the problem of explicitly
solving these equations.
In ref.~\cite{Kirillov}
an elegant algebraic formulation of the solution
for $\cal G$=A$_{n-1}$ and D$_{n-1}$
has however been given, based on the work of ref.~\cite{KR}
on the representations of Yangians.
For $\cal G$=A$_{n-1}$ the solution is especially simple
\begin{equation}
\xi_j^{(a)} = \frac{
\sin\!\left(\frac{a\pi}{n+\ell}\right)
\sin\!\left(\frac{(n-a)\pi}{n+\ell}\right)}{
\sin\!\left(\frac{(j+a)\pi}{n+\ell}\right)
\sin\!\left(\frac{(n+j-a)\pi}{n+\ell}\right)}
\end{equation}
and can in fact easily be checked by direct substitution in (\ref{TBAxi}).
For the exceptional cases no explicit form of the solutions
is known, but some conjectures towards a solution have been made
in ref.~\cite{Kuniba}.

Another remark to be made is that the identities (\ref{ADEdilog})
are in fact consequences of the stronger identities \cite{BR,Kirillov}
\begin{equation}
s^{\,\cal G}(\ell) \equiv  \frac{6}{\pi^2}
\sum_{a=1}^{\rG} \sum_{j=1}^{\ell}
L\left(\xi_j^{(a)}\right) = \frac{\ell \: \mbox{dim }{\cal G}}{\ell+g} \, ,
\end{equation}
where we recall that $p=\ell+g-1$ in (\ref{ccG}).

\nsection{Summary and discussion}\label{sec7}
In this paper we have presented conjectures for
polynomial identities of A-D-E type following
the `solvable lattice model'
approach of Melzer \cite{Melzer} (see also ref.~\cite{Berkovich}).
All polynomials identities
can be viewed as finitizations of Rogers--Ramanujan
type identities for branching functions associated with the
GKO pair ${\cal G}^{(1)} \oplus {\cal G}^{(1)} \supset {\cal G}^{(1)}$
at levels $\ell-1$, 1 and $\ell$, respectively.
Apart from extensive computer tests we have corroborated our conjectures
by studying the behaviour under level-rank duality
for $\cal G$=A$_{n-1}$ and by
establishing the expected dilogarithm identities following from
the A-D-E Rogers--Ramanujan in the asymptotic limit $q\to 1^{-}$.

\vspace*{5mm}

Given that we have no proofs of our conjectures a lot
of course remains to be done.
First of all it is to be noted that we have only presented
fermionic polynomials associated with one particular choice
of one-dimensional configuration sum, or, equivalently,
we have only given fermionic finitizations of one particular
branching function associated with the above mentioned coset pair.
This is in contrast to Melzer's original work where fermionic
polynomials for all Virasoro characters $\chi_{r,s}^{(p,p+1)}$
were conjectured.
In fact, the method of proof for the A$_1$ polynomial
identities \cite{BM} relies heavily on the completeness
of a set of fermionic polynomials in order to apply
recurrence methods similar to those used for obtaining the bosonic
polynomials.
However, despite some efforts to find
fermionic counterparts to all configuration sums
$X_L(a,b,c)$ we did not succeed
to do so in general.
So, for example, for $\cal G$=A$_{n-1}$ we
only managed to find fermionic forms corresponding
to configuration sums of the form
$X_L(\ell \Lambda_0,b,b+\e_j)$ with
$b=p\Lambda_{\mu}+(\ell-p)\Lambda_{\nu}$, $0\leq p \leq \ell$.
Even so, assuming that we would have been able to establish completeness,
it still seems that the recurrence method developed
in ref.~\cite{Berkovich} is far from ideal to tackle
the general A$_{n-1}$ case.
In this respect, exploiting higher rank partition theory
might be a more promising way to go.
(For recent progress on proofs of polynomial identities
for finitized Virasoro characters using
solely partition theoretic arguments, see ref.~\cite{FQ}.)

Setting aside the problem of proof, it is quite clear that
many of the results of this paper admit further generalization.
Instead of defining the fermionic polynomials (\ref{ADEpoly})
based on the constraint system (\ref{TBAconstr}),
we could have started with the more general equation
\begin{equation}
\m{j}{a} + \n{j}{a} =
\frac{1}{2} \left[ L \: \delta_{a,p} \delta_{j,s} +
\sum_{b=1}^{\rG}
{\cal I}^{\,\cal G}_{a,b}\:
\n{j}{b} +
\sum_{k=1}^{\ell-1} {\cal I}^{\A}_{j,k}
\m{k}{a} \right]
\label{TBAs}
\end{equation}
as considered in parameter dependent form in ref.~\cite{BR}.
The parameter $s$ herein provides the generalization
to the cosets
\begin{equation}
\begin{array}{rcccccc}
&{\cal G}^{(1)} &\oplus & {\cal G}^{(1)} &\supset & {\cal G}^{(1)} & \\
\mbox{level } &\ell-s& &  s& & \ell & ,
\label{GKOs}
\end{array}
\end{equation}
and by letting $p$ taking any of the values $1,\ldots,\mbox{rank }\cal G$,
instead of fixing it as in section~\ref{sec2},
we can obtain fermionic finitizations
to several and not just 1 branching function associated with
(\ref{GKOs}).
The problem is then  of course to also define appropriate
bosonic polynomials to match the fermionic expressions.
For the case $\cal G$=A$_1$, $s>1$ these should be provided
by CTM calculations of ref.~\cite{DJKMOb}
for the fused ABF models.

In ref.~\cite{Kuniba}
an even further generalization to (\ref{TBAconstr})
was proposed,
extending (\ref{TBAs}) to the case of non-simply laced Lie algebras.
Clearly, in defining the appropriately
generalized form
of the fermionic expression (\ref{ADEpoly}), this should for $s=1$
correspond
to the one-dimensional configuration sums for the B$_{n}^{(1)}$ and
C$_{n}^{(1)}$ RSOS models of ref.~\cite{DJKMO}.
In fact,
quite possibly the only change that has to be made in
(\ref{ADEpoly}) is the replacement of $\ell$ by $t_a \ell$
to account for the fact that not all simple roots $\alpha_a$
have equal length. Here $t_a$ is defined by normalizing
$|\mbox{long root}|^2=2$ and setting
$t_a=2/|\alpha_a|^2$, $a$ being the $a$-th simple root.

We hope to address some of the above mentioned problems and
generalizations in future publications.

\section*{Acknowledgements}
We acknowledge many helpful and stimulating discussions
on Rogers--Ramanujan identities with Omar Foda.
We also thank Omar Foda, Peter Forrester and Yu-kui
Zhou for kind interest and encouragements.
This work is supported by the Australian Research
Council.


\begin{thebibliography}{99}

\bibitem{Rogers}
L.~J.~Rogers,
{\em Proc.\ London Math.\ Soc.} {\bf 25} (1894) 318;
{\em Proc.\ Cambridge Phil.\ Soc.} {\bf 19} (1919) 211.

\bibitem{Ramanujan}
S.~Ramanujan,
{\em Proc.\ Cambridge Phil.\ Soc.} {\bf 19} (1919) 214.

\bibitem{Schur}
I.~J.~Schur,
{\em S.-B.\ Preuss.\ Akad.\ Wiss.\ Phys.-Math. Kl.}
(1917) 302.

\bibitem{Andrews}
G.~E.~Andrews,
{\em The Theory of Partitions}
(Addison-Wesley, Reading, Massachusetts, 1976).

\bibitem{Kac}
V.~G.~Kac, {\em Infinite dimensional Lie algebras}
(Birkh\"auser, Boston, 1983).

\bibitem{KP}
V.~G.~Kac and D.~H.~Peterson,
{\em Adv.\ Math.} {\bf 53} (1984) 125.

\bibitem{Rocha}
A.~Rocha-Caridi,
in {\em Vertex Operators in Mathematics and Physics},
eds.\ J.~Lepowsky, S.~Mandelstam and I.~M.~Singer
(Springer, Berlin, 1985).

\bibitem{BPZ}
A.~A.~Belavin, A.~M.~Polyakov and A.~B.~Zamolodchikov,
{\em J.\ Stat.\ Phys.} {\bf 34} (1984) 763;
{\em Nucl.\ Phys.} {\bf B241} (1984) 333.

\bibitem{Baxter}
R.~J.~Baxter,
{\em Exactly solved models in statistical mechanics}
(Academic Press, London, 1982).

\bibitem{ABF}
G.~E.~Andrews, R.~J.~Baxter and P.~J.~Forrester,
{\em J.\ Stat.\ Phys.} {\bf 35} (1984) 193.

\bibitem{FB}
P.~J.~Forrester and R.~J.~Baxter,
{\em J.\ Stat.\ Phys.} {\bf 35} (1985) 435.

\bibitem{KKMMa}
R.~Kedem, T.~R.~Klassen, B.~M.~McCoy and E.~Melzer,
{\em Phys.\ Lett.} {\bf 304B} (1993) 263.

\bibitem{KKMMb}
R.~Kedem, T.~R.~Klassen, B.~M.~McCoy and E.~Melzer,
{\em Phys.\ Lett.} {\bf 307B} (1993) 68.

\bibitem{Melzer}
E.~Melzer,
{\em Int.\ J.\ Mod.\ Phys.} {\bf A 9} (1994) 1115.

\bibitem{Berkovich}
A.~Berkovich,
Fermionic counting of RSOS-states and Virasoro
character formulas for the minimal series
$M(\nu,\nu+1)$. Exact results. Preprint BONN-HE-94-04, hep-th/9403073.
To appear in {\em Nucl.\ Phys.} {\bf B431} (1994).

\bibitem{GKO}
P.~Goddard, A.~Kent and D.~Olive,
{\em Phys.\ Lett.} {\bf 152B} (1985) 88.

\bibitem{BR}
V.~V.~Bazhanov and N.~{Yu}.~Reshetikhin,
{\em Int.\ J.\ Mod.\ Phys.} {\bf A 4} (1989) 115;
{\em J.\ Phys.\ A: Math.\ Gen.} {\bf 23} (1990) 1477;
{\em Prog.\ Theor.\ Phys.\ Suppl.} {\bf 102} (1990) 301.

\bibitem{JMOb}
M.~Jimbo, T.~Miwa and M.~Okado,
{\em Nucl.\ Phys.} {\bf B300 [FS22]} (1988) 74.

\bibitem{DJKMO}
E.~Date, M.~Jimbo, A.~Kuniba, T.~Miwa and M.~Okado,
{\em Lett.\ Math.\ Phys.} {\bf 17} (1989) 69.

\bibitem{WPSN}
S.~O.~Warnaar, P.~A.~Pearce, K.~A.~Seaton and B.~Nienhuis,
{\em J.\ Stat.\ Phys.} {\bf 74} (1994) 469.

\bibitem{Kuniba}
A.~Kuniba,
{\em Nucl.\ Phys.} {\bf B389} (1993) 209.

\bibitem{JMOa}
M.~Jimbo, T.~Miwa and M.~Okado,
{\em Commun.\ Math.\ Phys.} {\bf 116} (1988) 507.

\bibitem{Wolfram}
S.~Wolfram,
{\em Mathematica: A System for Doing Mathematics by Computer}
(Addison Wesley, Reading, Massachusetts, 1991).

\bibitem{unpublished}
S.~O.~Warnaar,
unpublished.

\bibitem{WP}
A sketch of the proof has been given in:
S.~O.~Warnaar and P.~A.~Pearce,
Exceptional structure of the dilute A$_3$ model:
E$_8$ and E$_7$ Rogers--Ramanujan identities.
Preprint Melbourne University, hep-th/9408136.
To appear in {\em J.\ Phys.\ A: Math.\ Gen.}

\bibitem{Zamolodchikov}
A.~B.~Zamolodchikov,
{\em Adv.\ Stud.\ in Pure Math.} {\bf 19} (1989) 1;
{\em Int.\ J.\ Mod.\ Phys.} {\bf A 4} (1989) 4235.

\bibitem{BNW}
V.~V.~Bazhanov, B.~Nienhuis and S.~O.~Warnaar,
{\em Phys.\ Lett.} {\bf 322B} (1994) 198.

\bibitem{BBSS}
F.~A.~Bais, P.~Bouwknegt, K.~Schoutens and M.~Surridge,
{\em Nucl.\ Phys.} {\bf B304} (1988) 371.

\bibitem{CR}
P.~Christe and F.~Ravanini,
{\em Int.\ J.\ Mod.\ Phys.} {\bf A 4} (1989) 897.

\bibitem{KN}
A.~Kuniba and T.~Nakanishi,
in {\em Modern Quantum Field Theory}, eds.
S.~Das, A.~Dhar, S.~Mukhi, A.~Raina and A.~Sen,
(World Scientific, Singapore, 1991).

\bibitem{RS}
B.~Richmond and G.~Szekeres,
{\em J.\ Austral.\ Math.\ Soc.\ (A)} {\bf 31} (1981) 362.

\bibitem{Lewin}
L.~Lewin,
{\em Polylogarithms and Associated Functions}
(Elsevier, Amsterdam, 1981).

\bibitem{KM}
T.~R.~Klassen and E.~Melzer,
{\em Nucl.\ Phys.} {\bf B338} (1990) 485.

\bibitem{AlB}
Al.~B.~Zamolodchikov,
{\em Nucl.\ Phys.} {\bf B342} (1990) 695.

\bibitem{Ravanini}
F.~Ravanini,
{\em Phys.\ Lett.} {\bf 282B} (1992) 73.

\bibitem{Kirillov}
A.~N.~Kirillov,
{\em J.\ Sov.\ Math.} {\bf 47} (1989) 2450.

\bibitem{NRT}
W.~Nahm, A.~Recknagel and M.~Terhoeven,
{\em Mod.\ Phys.\ Lett.} {\bf A 8} (1993) 1835.

\bibitem{DKKMM}
S.~Dasmahapatra, R.~Kedem, T.~R.~Klassen, B.~M.~McCoy and E.~Melzer,
{\em Int.\ J.\ Mod.\ Phys.} {\bf B 7} (1993) 3617.

\bibitem{Kirillovb}
A.~N.~Kirillov,
Dilogarithm identities.
preprint hep-th/9408113.

\bibitem{DV}
R.~Dijkgraaf and E.~Verlinde,
{\em Nucl.\ Phys.\ (Proc.\ Suppl.)} {\bf B5} (1988) 87.

\bibitem{KR}
A.~N.~Kirillov and N.~{Yu}.~Reshetikhin,
{\em J.\ Sov.\ Math.} {\bf 52} (1990) 3156.

\bibitem{BM}
For the characters $\chi_{r,1}^{(p,p+1)}$ the proof was given
in ref.~\cite{Berkovich}. A a proof for the
general $\chi_{r,s}^{(p,p+1)}$ case by
A.~Berkovich and B.~M.~McCoy has recently been announced.

\bibitem{FQ}
O.~Foda and Y.-H.~Quano,
Polynomial identities of the Rogers--Ramanujan type.
Preprint No. 25 Melbourne University (1994), hep-th/9407191.

\bibitem{DJKMOb}
E.~Date, M.~Jimbo, A.~Kuniba, T.~Miwa and M.~Okado,
{\em Nucl.\ Phys.} {\bf B290 [FS20]} (1987) 231.

\end{thebibliography}
\end{document}